\documentclass[useAMS,usenatbib,usegraphicx]{mn2e}
\title[Radio and molecular jet in G353.273$+$0.641]{Intermittent maser flare around the high mass young stellar object G353.273$+$0.641 II: Detection of a radio and molecular jet}
\author[K. Motogi et al.]{K. Motogi,$^{1}$\thanks{E-mail:motogi@yamaguchi-u.ac.jp}
K. Sorai,$^{2}$  K. Niinuma,$^{3}$ K. Sugiyama,$^{3}$ M. Honma,$^{4, 5}$ and K. Fujisawa$^{1, 3}$\\
$^{1}$The Research Institute for Time Studies, Yamaguchi University, Yoshida 1677-1, Yamaguchi, Yamaguchi, 753-8511, Japan\\
$^{2}$Department of Physics $/$ Department of Cosmosciences, Hokkaido University, N10 W8, Sapporo 060-0810, Japan\\
$^{3}$Department of Physics, Faculty of Science, Yamaguchi University, Yoshida 1677-1, Yamaguchi, Yamaguchi, 753-8512, Japan\\
$^{4}$Department of Astronomical Science, The Graduate University for Advanced Studies, 2-21-1 Osawa, Mitaka, Tokyo 181-8588, Japan\\
$^{5}$Mizusawa VLBI Observatory, National Astronomical Observatory of Japan, 2-12 Hoshi-ga-oka, Mizusawa, Oshu, Iwate 023-0861, Japan\\}
\begin{document}

\date{Accepted 2012 September 18. Received 2012 September 17; in original form 2012 August 28}

\pagerange{\pageref{firstpage}--\pageref{lastpage}} \pubyear{2012}

\maketitle

\label{firstpage}

\begin{abstract}
We report the first detection of a radio-continuum and molecular jet associated with a dominant blue-shifted maser source, G353.273+0.641. 
A radio jet is extended 3000 au along NW-SE direction. H$_{2}$O masers are found to be clustered in the root of a bipolar radio jet. 
A molecular jet is detected by thermal SiO ($\upsilon$ = 0, $J$ = 2-1) emission. The SiO spectrum is extremely wide (-120 -- +87 km s$^{-1}$) and significantly blue-shift dominated, similar to the maser emission. The observed geometry and remarkable spectral similarity between H$_{2}$O maser and SiO strongly suggests the existence of a maser-scale ($\sim$ 340 au) molecular jet that is enclosed by the extended radio jet. We propose a “disc-masking” scenario as the origin of the strong blue-shift dominance, where an optically thick disc obscures a red-shifted lobe of a compact jet. 
\end{abstract}


\section{Introduction}
The dominant blue-shifted maser (DBSM) is a class of 22 GHz H$_{2}$O maser sources that shows a highly asymmetric spectrum, 
where almost all flux is concentrated in blue-shifted emission, with a broad velocity range (up to $\pm$ 100 km s$^{-1}$) with respect to the systemic velocity (\citealt{Caswell2008}; hereafter CP08). 
It is a candidate of high mass protostellar object (HMPO) with a pole-on jet, and often observed in the earliest evolutional phase accompanied by the class II CH$_{3}$OH masers \citep{Breen2010}. 

The target source G353.273+0.641 (hereafter G353) is the archetypal DBSM at a distance of 1.7 kpc (CP08). 
Motogi et al. (2011) (hereafter MO11) have found highly intermittent maser flare activities in G353 based on a long-term VLBI $+$ single-dish monitoring studies. 
Observed maser flares were accompanied by systematic changes of a maser distribution, indicating recurrent shock propagations. 
Therefore, MO11 have concluded that the time-dependent maser activities are caused by an episodic high mass protostellar jet. 

In this letter, we report new detections of a radio-continuum and molecular jet in G353 with the Australia Telescope Compact Array (ATCA) and the Nobeyama Radio Observatory (NRO) 45m telescope. 
An additional VLBI dataset of the H$_{2}$O maser taken by the VLBI Exploration of Radio Astrometry (VERA) is also used for a comparison between the radio jet and the maser. 
These new observations intend to confirm whether G353 is really excited by a jet or not. 
This is the first detection of a host protostellar jet associated with a known DBSM, supporting a previous "pole-on jet" hypothesis in CP08 and MO11. 

\section{Observations and data reduction}
\subsection{ATCA}
The ATCA radio continuum observation was made on 2012, Jan. 12 with the array configuration of 6A. 
Since there was a receiver trouble in CA05, totally five antennae were used for 10 hours synthesis. 
We observed two 2 GHz bands centred on 18 and 22 GHz simultaneously, using the Compact Array Broadband Backend (Wilson et al. 2011) in the 2048 channel mode. 
The H$_{2}$O maser emission is also contained in the 22 GHz band. 
The phase centre of the array was ($\alpha$,$\delta$)$_{\rmn{J2000.0}}$ = ($17^{\rmn{h}}26^{\rmn{m}}01^{\rmn{s}}.59$, -34\degr 15\arcmin 10\arcsec.00). 
This is slightly ($\sim 5\arcsec$) north from the maser position reported in MO11, 
in order to avoid an effect of any artefact that can be caused by instrumental errors. 
Observation was done in the fast-switching mode using the nearby bright ($\sim$ 1 Jy) calibrator 1713-336 (1.8$\degr$ from G353). 
The cycle time was set to $\sim$ 200 sec, where we spent 150 sec for G353 and 50 sec for 1713-336 including slew time. 
The pointing accuracy was checked every hour using 1713-336. 
 
Data reduction was performed with the MIRIAD package.   
Bandpass calibration was performed using 1921-293. 
The primary flux calibrator 1934-638 was used for absolute flux calibration. 
Estimated fluxes of 1.10 Jy (18 GHz) and 0.83 Jy (22 GHz) are well consistent with the previous records in the ATCA Calibrator Database, 
and hence, we conclude that our flux calibration is better than nominal 10 \% accuracy. 
After the standard calibration procedures, we successfully imaged a radio jet in G353 with a robust weighting (robust = 0.5 in the task "invert"). 
The synthesised beam sizes (full width half maximum : FWHM) are 0\arcsec.84 $\times$ 0\arcsec.45 and 0\arcsec.69 $\times$ 0\arcsec.36 at 18 and 22 GHz, respectively, with a beam position angle of 8.7\degr. 

We finally performed a self-calibration to maximise image dynamic ranges.  
Since the target source is too weak to be detected in individual scan, all target scans were integrated over, in order to correct a residual gain offset between each antennas. 
This increased image dynamic range by about 16 and 8\% for 18 and 22 GHz, respectively. 
The 1-$\sigma$ noise levels achieved in final images are 19 and 37 $\mu$Jy beam$^{-1}$ at 18 and 22 GHz, respectively. 
We also imaged an H$_{2}$O maser spot at the intensity peak, which was strong enough to be detected in our coarse spectral resolution ($\sim$ 13 km s$^{-1}$). 
It was used for an evaluation of our astrometric accuracy by compariing the maser position derived from a new VLBI dataset (see section 2.3 and 3.1). 

\subsection{NRO 45m}
High-sensitivity single-dish observations with the NRO 45m telescope were carried out on 2012, Mar. 12 -- 14 and Apr. 16 -- 17. 
We observed SiO ($\upsilon$ = 0, $J$ = 2-1) transition at 86.84696 GHz in a single pointing toward G353, searching for a molecular jet. 
We also conducted mapping observation of the natal molecular core and it will be reported in a forthcoming paper (Motogi et al. 2012 in prep). 

We used a waveguide-type dual-polarisation sideband-separating SIS receivers, T100V/H \citep{Nakajima2010}. 
The beam size (FWHM) is $\sim$ 19\arcsec (19\arcsec.2 for T100V and 18\arcsec.9 for T100H), corresponding a linear scale of 0.16 pc at 1.7 kpc.  
The backend system was the Spectral Analysis Machine for the 45m telescope (SAM45: e.g., \citealt{Iono2012}). 
It provides total 16-array outputs (e.g., 2 polarisations $\times$ 2 sidebands $\times$ 4 spectral windows), each with 4096 spectral channels. 
A bandwidth per array is selectable between 16 MHz -- 2 GHz, and we adopted 1-GHz bandwidth mode (244.14 kHz per spectral channel). 

The observations were made in the position-switching method. 
The absolute pointing accuracy was checked every hour with a strong SiO maser source using a 43GHz SIS receiver (S40). 
It was within $\pm$ 4\arcsec throughout the observations. 
The system noise temperature ($T_{\rm sys}$) was typically 200 - 300 K in single sideband. 
Intensity scale was calibrated by the standard chopper-wheel method. 

Data were reduced with the NEWSTAR software package developed at NRO. 
Observed antenna temperature was converted into main beam brightness temperature ($T_{\rm mb}$) scale, 
assuming the main beam efficiency of 40\% and 43\% for T100V and T100H, respectively. 
Two distinct polarisations were averaged in $T_{\rm mb}$ scale for sensitivity.  
The initial 4096 channels were smoothed every eight channels, resulting in the final spectral resolution of 6.8 km s$^{-1}$. 
Total integration time was about 1 hour per polarisation and r.m.s noise level of 6.4 mK in $T_{\rm mb}$ scale was achieved. 

\subsection{VERA}
New VLBI observation of the H$_{2}$O maser was done on 2012, Jan. 13 (a day after the ATCA observation) using the VERA.  
Typical $T_{\rm sys}$ was 300 -- 500 K in 6-hours observation, except for initial 2-hours where we observed rather high $T_{\rm sys}$ of 1000 K at the Ishigaki station. 
Detailed observing method and data reduction procedures has already been described in MO11. 
The position of the peak maser spot (-60.7 km s$^{-1}$) is determined with the astrometric accuracy of 0.10 and 0.34 milli-arcseconds (mas) in RA and DEC, respectively. 

\section{Results}
\subsection{Radio continuum:  jet plus disc?}
Figure 1 shows contour images of the radio continuum at 18 and 22 GHz, 
revealing elongation along the NW-SE direction with a total extent of $\sim$ 3000 au (3$\sigma$-threshold). 
Estimated flux and spectral index $\alpha$, between these frequencies are listed in Table 1. 
Although it still contains large error, the $\alpha$ that is estimated from the total fluxes is clearly positive. 
This can be a signature of an optically thick free-free emission, which is a typical case for a radio jet in both of high and low mass star-forming regions (e.g., \citealt{Anglada1996}). 
However, because of the limited angular resolution, we cannot exclude any sub-structure of the apparent radio lobes at present. 
We interpret the emission as a jet, but also consider the possibility of an optically thick disc around the host HMPO (see section 4).

\begin{figure*}
\includegraphics*[trim=0 0 0 0,scale=0.5]{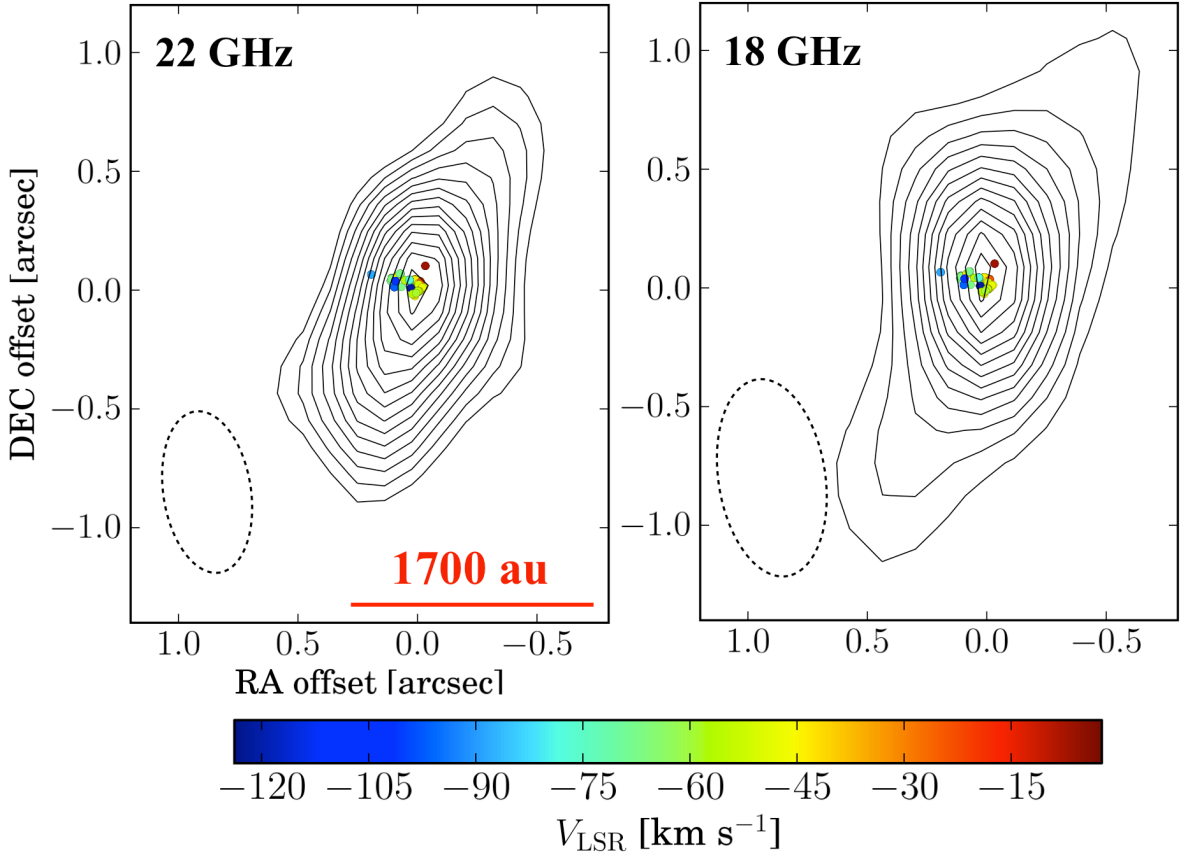}\\
\caption{Contour images of the radio jet in G353. 
Both contours are in steps of 37 $\mu$Jy beam$^{-1}$ with the minimum level of 3-$\sigma$ (56 and 111 $\mu$Jy beam$^{-1}$ at 18 and 22 GHz, respectively). 
Coordinate origin is ($\alpha$,$\delta$)$_{\rmn{J2000.0}}$ = ($17^{\rmn{h}}26^{\rmn{m}}01^{\rmn{s}}.5883$, -34\degr 15\arcmin 14\arcsec.905), that is the reference position in MO11. 
A synthesised beam is shown in the lower left corner of each panel. 
Colour points show all H$_{2}$O maser spots detected in MO11 with its line-of-sight (LOS) velocity ($V_{\rmn{LSR}}$; see colour scale at the bottom). }
\end{figure*}

The spatial relation between the maser and the radio continuum is also shown in Figure 1, where all maser spots detected in MO11 are superposed. 
As mentioned above, we directly checked our absolute astrometric accuracy at 22 GHz using the H$_{2}$O maser.  
Table 2 contains the positions of the peak maser spot measured in ATCA and VERA. 
The positional offset suggests that our astrometric accuracy is better than 30 mas at 22 GHz. 
We also directly evaluated the accuracy at 18 GHz from the r.m.s phase ($\phi_{\rmn{r.m.s}}$) of cross power spectrum. 
Estimated $\phi_{\rmn{r.m.s}}$ is $\sim$ 25\degr, i.e., 7\% of the fringe spacing, and hence, we conclude that the positional uncertainty at 18 GHz is better than $\sim$ 60 mas (7\% of the synthesised beam). 

The compact ($\sim$ 100 mas) maser cluster is clearly located at the root of the bipolar jet. 
The close location of the masers and continuum peaks implies that they have the same driving source, which is probably located within a few hundred mas. 
This kind of geometry is observed in several HMPOs and can naturally explain the high variability of the maser reported in MO11. 
In some cases, a maser seems to trace a rotating disc (e.g., $IRAS$16547-4247; \citealt{Franco2009}), but this cannot be the case in G353, since the masers velocities are too high to be explained by a disc rotation.

\begin{table}
  \caption{Properties of the radio jet}
   \begin{tabular}{ccc}\hline
   Frequency & Peak Flux & Total Flux \\
      (GHz)  &  (mJy beam$^{-1}$) & (mJy)  \\ \hline
18 & 0.52  $\pm$ 0.02 & 0.83 $\pm$ 0.08$^{\rmn{a}}$ \\
22 &  0.65 $\pm$ 0.04 & 1.41 $\pm$ 0.14$^{\rmn{a}}$ \\ \hline
Spectral Index$^{\rmn{b}}$ & \multicolumn{2}{c}{2.6 $\pm$ 1.0} \\ \hline
\multicolumn {3} {l} {$^{a}$ Nominal 10\% error is adopted (see section 2.1).}\\
\multicolumn {3} {l} {$^{b}$ The index was calculated from the total flux.}\\
\end{tabular}
\end{table}

\begin{table}
  \caption{The positions of the peak maser spot}
   \begin{tabular}{ccc}\hline
   Array & RA$_{\rmn{J2000.0}}$ & DEC$_{\rmn{J2000.0}}$ \\ \hline
ATCA & $17^{\rmn{h}}26^{\rmn{m}}01^{\rmn{s}}.5885 (4)^{\rmn{a}}$ & -34\degr 15\arcmin 14\arcsec.908 (9)\\
VERA & $17^{\rmn{h}}26^{\rmn{m}}01^{\rmn{s}}.588469 (7)$ & -34\degr 15\arcmin 14\arcsec.8900 (3) \\ \hline
Offset  &  +7 mas & -27 mas \\ \hline
\multicolumn {3} {l} {$^{a}$ All numbers in parentheses represent the errors in }\\
\multicolumn {3} {l} {units of the last significant digit.}\\
\end{tabular}
\end{table}

\subsection{Molecular jet}
Figure 2 shows the detected extremely high-velocity (EHV) SiO spectrum. 
We detected an unusually flat and continuous blue-shifted emission (6$\sigma$) and isolated red-shifted emission at +87 km s$^{-1}$ (5$\sigma$). 
The latter is identified as the SiO emission because its frequency is not matched with a rest frequency of any known molecular transition, and also, 
it is within the range of Doppler-shift reported for the H$_{2}$O maser emission (see below). 
The total velocity range (full width at zero intensity: FWZI) of 210 km s$^{-1}$ is one of the extreme cases compared to the previous SiO (2-1) survey toward high mass star-forming regions (e.g., \citealt{Lopez2011}). 

\begin{figure*}
\includegraphics*[trim=0 0 0 0,scale=0.6]{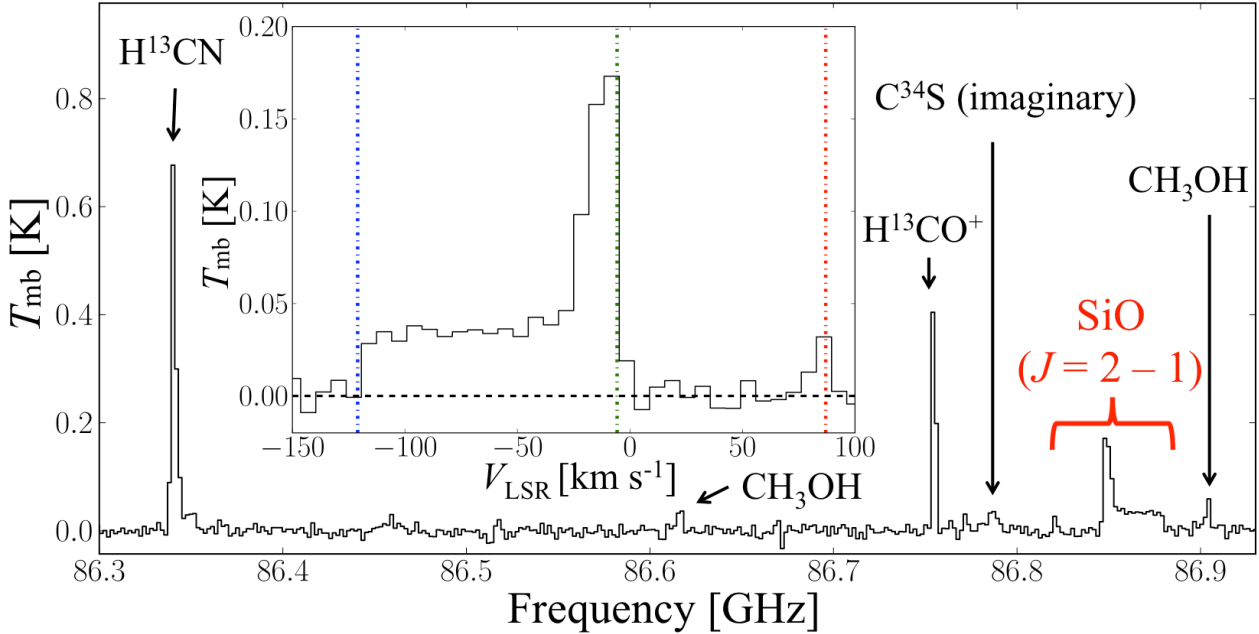}\\
\caption{A part of the 1GHz-wide spectrum taken by NRO 45m telescope. 
We marked all identified lines including the targeted SiO ($\upsilon$ = 0, $J$ = 2-1) transition. 
The subplot shows the magnified spectrum of the SiO line. 
The blue and red dotted line indicates the most red (+87 km s$^{-1}$) and blue-shifted (-120 km s$^{-1}$) H$_{2}$O maser component reported in MO11 and CP08, respectively . 
The systemic velocity of -5 km s$^{-1}$ is also indicated by the green dotted line. 
The black horizontal line shows zero level determined by a linear baseline subtraction. 
The apparent emission at 86.78 GHz is a reflected image of C$^{34}$S ($J$ = 2-1) line at 96.4129495 GHz in the upper sideband. }
\end{figure*}

Surprisingly, observed spectral features of the EHV SiO jet are very similar to that of the H$_{2}$O maser in CP08 and MO11, 
i.e., very broad velocity range on the blue-shifted side (-120 -- -5 km s$^{-1}$), no emission on the red-shifted side (0 -- $+$80 km s$^{-1}$) and isolated emission at $+$87 km s$^{-1}$. 
We note that the narrow peak at the systemic velocity seems to be an ambient shock or extended low-velocity outflow, 
since it was detected in all mapping points (4$\times$4 Nyquist grid of 9\arcsec) of the NRO observation (Motogi et al. 2012 in prep). 
It is probably independent of the SiO jet, that is only detected at the position of the maser. 

The remarkable spectral similarity strongly suggests that the SiO jet is directly related to the maser excitation. 
If this is the case, a scale of the SiO jet can be comparable with the maser distribution, i.e., only 200 mas ($\sim$ 340 au). 
A dilution-corrected brightness temperature of EHV component is $\sim$ 550 K in this case. 
This is consistent with a typical kinetic temperature $T_{\rmn{kin}}$ of a SiO jet (e.g., \citealt{Lee2010}), indicating that the jet is almost thermalised. 

\section{Discussion: Origin of the blue-shift dominated jet}
The detection of the highly blue-shifted SiO jet manifests that the fundamental origin of the blue-shift dominance is not a matter of the radiative transfer of H$_{2}$O masers that was discussed in CP08. 
The most simple explanation for the observed blue-shift dominance in the maser and SiO emission is a highly asymmetric jet structure. 
However, this is an unlikely case, since the observed radio jet shows a clear bipolar structure. 
One important fact reported in \citet{Breen2010} is that the number of DBSMs is significantly larger than that of a red-shift dominated source ($<$ 20 - 30 \%; see also the data in \citealt{Urquhart2011}). 
Such a statistical anomaly requires some intrinsic mechanism to selectively cause blue-shift dominance. 

Instead of an asymmetric jet, we propose a "disc-masking" scenario, where red-shifted emission is masked by a disc which is optically thick at 22 and 86 GHz. 
Observational signatures of both optically thick free-free and dusty disc are reported for several sources. 
The most famous example of an ionized disc is seen in the source I in Orion KL (\citealt{Reid2007}), where the compact ionized disc seems to be optically thick up to 86 GHz (e.g., \citealt{Beuther2004}). 
On the other hand, a dusty disc which is optically thick at 86 GHz is often detected in low mass class 0 objects (e.g., \citealt{Chandler2005}; \citealt{Hirano2010}), and sometimes, in a high mass object (e.g., \citealt{Vandertak2005}). Typical size of these optically thick discs ($\sim$ a few hundred astronomical units) are actually comparable with the maser-scale molecular jet mentioned above. 

Our scenario is a natural explanation from the view point of a core collapse and disc-accretion, i.e., a younger object has a more massive and dense accretion disc. 
It can consistently explain why DBSMs are observed in the youngest evolutional phase of HMPOs (e.g., \citealt{Breen2010}). 
Expected pole-on geometry of DBSM is also suitable for the scenario, for example, 
the inclination angle of the jet in G353 estimated from the LOS velocity and maser proper motion is smaller than 30$^{\circ}$ (see MO11). 
This pole-on geometry can allow a disc to mask significant fraction of the red-shifted jet, 
although the detection of highly red-shifted component of +87 km s$^{-1}$ suggests that the masking effect is incomplete, possibly because of a relatively small disc size compared to the jet. 

Figure 3 shows a schematic view of the expected source geometry. 
If our scenario is true, the optically thick 22 GHz emission detected by the ATCA should be a combination of an unresolved optically thick disc and an extended radio jet. 
The apparent geometry, where blue-shifted masers are enclosed by the radio jet, may require that the radio jet itself is optically thin at 22 GHz. 
A direct imaging of the unusual SiO jet and cm -- mm continuum in G353 with a 100-mas resolution is absolutely necessary to verify our scenario. 
It will be a first step of further statistical discussion about exact natures of DBSMs. 
A proper motion measurement of the jet is also important to understand the relation between the ionized and molecular component of the jet.

\begin{figure}
\centering
\includegraphics*[scale=0.6]{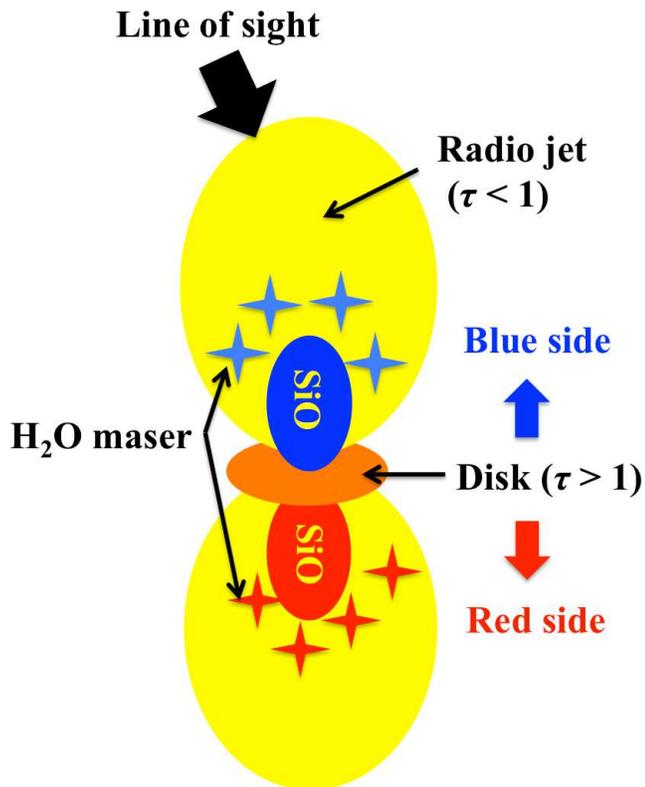}
\caption{Schematic view of the source geometry in our "disc-masking" scenario.}
\end{figure}

\section*{Acknowledgments}
The authors thank the staff members of the ATCA, NRO and VERA for their generous assistance in the observations and data reductions. 
The ATCA is part of the Australia Telescope National Facility which is funded by the Commonwealth of Australia for operation as a National Facility managed by CSIRO. 
The NRO is a branch of National Astronomical Observatory of Japan. 
The VERA is operated by the Mizusawa VLBI observatory that is also a branch of National Astronomical Observatory of Japan. 
This work was financially supported by the Grant-in-Aid for the Japan Society for the Promotion of Science Fellows (K. M.).

\label{lastpage}

\end{document}